\documentclass[notitlepage,reprint, aps, prl, twocolumn,superscriptaddress,nofootinbib,floatfix,secnumarabic]{revtex4-2}

\usepackage[normalem]{ulem}
\usepackage{amsmath,amssymb,amsthm,mathtools}
\usepackage{graphicx}
\usepackage{bm}
\usepackage[dvipsnames]{xcolor}
\usepackage{physics}
\usepackage{booktabs}
\usepackage{longtable}

\usepackage{comment}
\usepackage{lipsum}

\usepackage{txfonts}
\usepackage{graphicx}     
\usepackage{dcolumn}      
\usepackage{bm}           
\usepackage{mathtools,amsmath,amssymb}
\usepackage{physics}

\usepackage{tabularray}

\usepackage{booktabs}

\usepackage{comment}

\usepackage{quantikz}
\usepackage{adjustbox}

\begin{document}
\raggedbottom
\setcounter{secnumdepth}{3}
\renewcommand{\thesection}{\arabic{section}}

\title{Variational Quantum-Neural Hybrid Imaginary-Time Evolution}

\author{Hiroki Kuji}
\email{1235702@ed.tus.ac.jp}
\affiliation{Department of Physics, Tokyo University of Science, Shinjuku, Tokyo 162-8601, Japan}
\affiliation{Department of Electrical, Electronic, and Communication Engineering, Faculty of Science and Engineering, Chuo University, Bunkyo-ku, Tokyo 112-8551, Japan}
\author{Tetsuro Nikuni}
\affiliation{Department of Physics, Tokyo University of Science, Shinjuku, Tokyo 162-8601, Japan}
\author{Yuta Shingu}
\affiliation{Department of Physics, Tokyo University of Science, Shinjuku, Tokyo 162-8601, Japan}

\begin{abstract}
Numerous methodologies have been proposed to implement imaginary-time evolution (ITE) on quantum computers. Among these, variational ITE (VITE) methods for noisy intermediate-scale quantum
(NISQ) computers have attracted much attention, as they use parameterized quantum circuits to mimic non-unitary dynamics.
Although widely studied, conventional variational quantum algorithms face challenges in achieving high accuracy due to their strong dependence on the choice of ansatz quantum circuits. 
Recently, the variational quantum-neural hybrid eigensolver (VQNHE), which combines neural networks (NN) with a variational quantum eigensolver, has been proposed. This approach enhances the expressive power of variational states and improves the estimation of expectation values. Motivated by this idea, we explore the hybridization of VITE with an NN-based non-unitary operator. In this study, we propose a method named variational quantum-neural hybrid
ITE (VQNHITE). By combining the NN with a parameterized quantum circuit, our proposal enhances the expressive power compared to conventional approaches, enabling more accurate tracking of imaginary-time dynamics. In addition, to mitigate the instability arising from randomly initialized NN parameters, we introduce an initial-parameter optimization procedure at a small imaginary-time step, which stabilizes the subsequent variational evolution. We tested our approach with numerical simulations on Heisenberg spin chains under both nearest-neighbor and all-to-all circuit
connectivities. The results demonstrate that VQNHITE consistently achieves higher fidelity with the exact ITE state compared to VITE.
\end{abstract}
\maketitle

%%%%%%%%%%%%%%%%%%%%%%%%%%%%%%%%%%%%%%%%%%%%%%%%%%%%%%%%%%%
%%%%%%%%%%%%%%%%%%%%%%%%%%%%%%%%%%%%%%%%%%%%%%%%%%%%%%%%%%%
\section{Introduction}
%%%%%%%%%%%%%%%%%%%%%%%%%%%%%%%%%%%%%%%%%%%%%%%%%%%%%%%%%%%
%%%%%%%%%%%%%%%%%%%%%%%%%%%%%%%%%%%%%%%%%%%%%%%%%%%%%%%%%%%
Imaginary-time evolution (ITE) is the dynamics obtained by replacing the real-time parameter $t$ with $t\rightarrow-i\beta$ using a real parameter $\beta$ in the Schr\"{o}dinger equation, resulting in the time evolution operator $e^{-\hat{H}\beta}$, where $\hat{H}$ is the Hamiltonian.
ITE is a powerful approach that leads to numerous applications, such as ground-state search~\cite{lin2021real, schuch2007computational} and the preparation of thermal equilibrium states~\cite{mcardle2019variational} in a wide range of fields, including condensed matter physics, high-energy physics, and quantum chemistry~\cite{houck2012chip,horikiri2016high,o2016scalable,gerritsma2010quantum,cirac2012goals,georgescu2014quantum,byrnes2010mott,buluta2009quantum,foulkes2001quantum}.

However, the operator $e^{-\hat{H}\beta}$ is non-unitary and cannot be directly realized on quantum devices.
To address this challenge, various approaches have been developed to approximate ITE with quantum computers, such as variational ITE (VITE)~\cite{jones2019variational,mcardle2019variational,gacon2024variational}, quantum ITE~\cite{yeter2022quantum, yeter2021scattering,sun2021quantum,yeter2020practical}, and probabilistic ITE~\cite{lin2021real,liu2021probabilistic} algorithms. Among these methods, the VITE algorithm is particularly suitable for noisy intermediate-scale quantum (NISQ) devices~\cite{Preskill2018quantumcomputingin}, as it approximates the ITE dynamics using a parameterized quantum circuit updated through a hybrid quantum-classical loop. Despite its practicality, VITE suffers from a well-known limitation: its accuracy strongly depends on the expressive power of the choice of the ansatz, and insufficient expressivity can significantly degrade the fidelity of the evolved state. 

Another representative application of NISQ devices is the variational quantum eigensolver (VQE)~\cite{peruzzo2014variational, mcclean2016theory,o2016scalable,hempel2018quantum,liu2019variational, cao2019quantum, mcardle2020quantum, bauer2020quantum,amaro2022filtering}, which estimates the ground-state energy with a parameterized circuit $\hat{U}(\bm{\theta})$. Just as in VITE, the quality of the result is constrained by the choice of ansatz. To overcome this bottleneck, the variational quantum-neural hybrid eigensolver (VQNHE)~\cite{zhang2022variational} was recently proposed. VQNHE
enhances the expressivity by introducing a non-unitary operator $\hat{f}(\bm{\phi})$ generated by a classical neural network (NN), enabling improved accuracy in variational ground-state energy estimation. Similar NN-assisted quantum variational schemes have also been explored to enhance the representational power of quantum states~\cite{liu2018differentiable, liu2021solving, verdon2019quantum, hsieh2021unitary, benedetti2021variational, rivera2021avoiding,torlai2020precise,bennewitz2022neural,sagingalieva2023hybrid,liu2023training,miao2024neural}. Expectation values can be evaluated from a state vector modified by the non-unitary operator through post-processing.

\begin{figure*}
    \centering
    \includegraphics[width=0.8\linewidth]{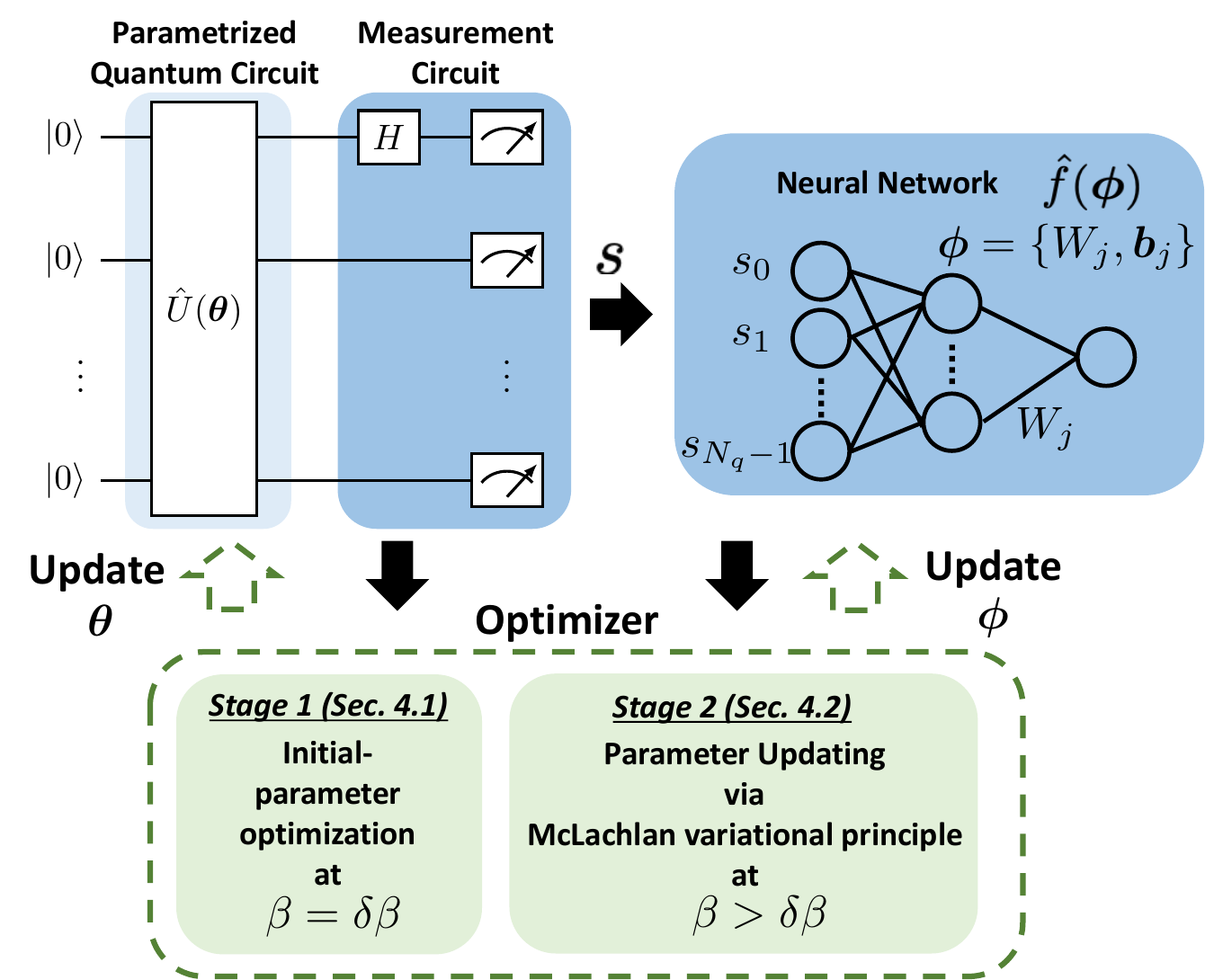}
    \caption{(Color online) Schematic workflow of VQNHITE. The parameterized quantum circuit is followed by a measurement circuit, and computational-basis measurements produce an outcome $s$. The measurement outcome is used in the neural-network post-processing to evaluate $f_{\bm{\phi}}$. 
    The lower-left panel represents the initialization step at $\beta=\delta \beta$, where the parameters $\bm{\lambda}=(\bm{\theta},\bm{\phi})$ are classically optimized using the cost function $F_{\mathrm{cost}}$. The lower-right panel represents the iterative VQNHITE step for $\beta > \delta \beta$, where the hybrid variational quantities are evaluated and the combined parameters are updated via the McLachlan variational principle.}
    \label{fig:scheme}
\end{figure*}

Motivated by these developments, we propose and investigate a VQNHITE approach that incorporates the NN-based non-unitary transformation into variational ITE, as shown in Fig.~\ref{fig:scheme}. Specifically, we construct the trial quantum state $\ket*{\tilde{\varphi}(\bm{\theta}, \bm{\phi})}=\hat{f}(\bm{\phi})\hat{U}(\bm{\theta})\ket*{\bar{0}}$,
where $\ket*{\bar{0}}$ is an initial state, $\hat{U}(\bm{\theta})$ is a unitary operator implemented by a parameterized quantum circuit, and $\hat{f}(\bm{\phi})$ is a non-unitary operator generated by the neural network. By combining these two ingredients, VQNHITE provides greater expressivity than standard VITE, which leads to a more accurate approximation of imaginary-time dynamics. In addition, to mitigate the sensitivity of NN-based methods to initial parameters, we adopt a gradient-based procedure to initialize the NN and circuit parameters, thereby improving both convergence and fidelity. We verify the performance of VQNHITE by numerical
simulations and compare it with conventional VITE using the fidelity
between variationally evolved states and the exact ITE state. Our results demonstrate that VQNHITE consistently achieves a higher fidelity than VITE. These
observations indicate that the NN-assisted non-unitary layer substantially mitigates the expressivity limitations inherent in standard circuit-based variational ITE.

This paper is organized as follows. In Sect.~\ref{sec:VITE}, we present an overview of VITE. In Sect.~\ref{sec:VQNHE}, we briefly review VQNHE. In Sect.~\ref{sec:VQNHITE}, we introduce our VQNHITE framework, detailing the combination of the variational quantum circuit and NN for ITE. In Sect.~\ref{sec:result}, we present numerical demonstrations comparing VQNHITE and VITE. Finally, in Sect.~\ref{sec:discussion}, we discuss and summarize our findings.

%%%%%%%%%%%%%%%%%%%%%%%%%%%%%%%%%%%%%%%%%%%%%%%%%%%%%%%%%%%
%%%%%%%%%%%%%%%%%%%%%%%%%%%%%%%%%%%%%%%%%%%%%%%%%%%%%%%%%%%
\section{Overview of Variational Imaginary-Time Evolution}\label{sec:VITE}
%%%%%%%%%%%%%%%%%%%%%%%%%%%%%%%%%%%%%%%%%%%%%%%%%%%%%%%%%%%
%%%%%%%%%%%%%%%%%%%%%%%%%%%%%%%%%%%%%%%%%%%%%%%%%%%%%%%%%%%
In this section, we provide an overview of the VITE method, which is suitable for NISQ devices. ITE is governed by the Wick-rotated Schr\"{o}dinger equation~\cite{PhysRev.96.1124, peskin2018introduction}, which can be expressed as
\begin{align}
    \frac{d \ket*{\psi(\beta)}}{d\beta}=-(\hat{H}-\bra*{\psi(\beta)}\hat{H}\ket*{\psi(\beta)}) \ket*{\psi(\beta)},
    \label{eq:wickrotated}
\end{align}
where $\hat{H}$ is a Hamiltonian and the expectation value $\expval{\hat{H}}{\psi(\beta)}$ ensures that $\ket*{\psi(\beta)}$ remains normalized. Accordingly, the state at imaginary time $\beta$ is given by
\begin{align}
    \ket*{\psi(\beta)}= \frac{\mathrm{exp}(-\hat{H} \beta) \ket*{\psi(0)}}{\sqrt{\bra*{\psi(0)}\mathrm{exp}(-2 \hat{H}\beta) \ket*{\psi(0)}}},
    \label{eq:state of wickrotated}
\end{align}
which has unit norm.

Since the imaginary-time operator $\mathrm{exp}(-\hat{H} \beta)$ is non-unitary, we cannot directly prepare the state in Eq.~\eqref{eq:state of wickrotated} on quantum devices. Instead, the variational imaginary-time simulation algorithm utilizes a parameterized wavefunction $\ket*{\varphi (\bm{\theta}(\beta))}$, implemented by a parameterized quantum circuit:
\begin{align}
    \ket*{\varphi (\bm{\theta}(\beta))}&= \hat{U}(\bm{\theta}(\beta))\ket*{\bar{0}} \nonumber \\
    \hat{U}(\bm{\theta}(\beta))& =\hat{U}_{N_{\mathrm{ p}}}(\theta_{N_{\mathrm{ p}}})\dotsm \hat{U}_{2}(\theta_{2})\hat{U}_{1}(\theta_{1}),
\end{align}
where $\hat{U}_i(\theta_i)$ $(i=1,\dotsc, N_{\mathrm{p}})$ represents a parameterized gate, $N_{\mathrm{p}}$ is the total number of parameters, $\theta_i$ is a real parameter, and $\ket*{\bar{0}}$ is the initial state of the ansatz, typically chosen as $\ket*{\psi(0)}$.

Next, we derive the evolution of the parameters from Eq.~\eqref{eq:wickrotated}. To obtain $\partial\bm{\theta}(\beta)/\partial\beta$, we apply the McLachlan variational principle~\cite{mcardle2019variational}, which minimizes the distance:
\begin{align}
    \delta \left\|\left(\frac{d}{d \beta} + \hat{H}-\bra*{\varphi (\bm{\theta}(\beta))}\hat{H}\ket*{\varphi (\bm{\theta}(\beta))}\right)\ket*{\varphi (\bm{\theta}(\beta))}  \right\|=0,
\end{align}
where $||\ket*{\psi}||\equiv\sqrt{\braket{\psi}}$ represents the norm of $\ket*{\psi}$.
This leads to the variational update equation
\begin{align}
    M\frac{d \bm{\theta}(\beta)}{d \beta}=\bm{C},
    \label{eq:MC}
\end{align}
where
\begin{align}
    M_{j,k}&=\Re\left( \frac{\partial \bra*{\varphi(\bm{\theta}(\beta))}}{\partial \theta_j}\frac{\partial \ket*{\varphi(\bm{\theta}(\beta))}}{\partial \theta_k} \right),\
    \label{Eq:Mk,j}\\
    C_j&= - \Re \left(\bra*{\varphi(\bm{\theta}(\beta))}\hat{H} \frac{\partial \ket*{\varphi(\bm{\theta}(\beta))}}{\partial \theta_j} \right).
\end{align}
The matrix $M$ and the vector $\bm{C}$ have sizes determined by the number of parameters $N_{\mathrm{ p}}$ rather than the number of qubits. Equation~(\ref{eq:MC}) can be approximated using the Euler method with a small step size $\delta \beta$, yielding
\begin{align}
    \bm{\theta}(\beta+\delta \beta)\simeq \bm{\theta}(\beta)+ M^{-1}(\beta)\cdot \bm{C}(\beta)\delta \beta.
    \label{eq:renew_param}
\end{align}

\begin{figure}[t]
    \centering
    \includegraphics[width=\columnwidth]{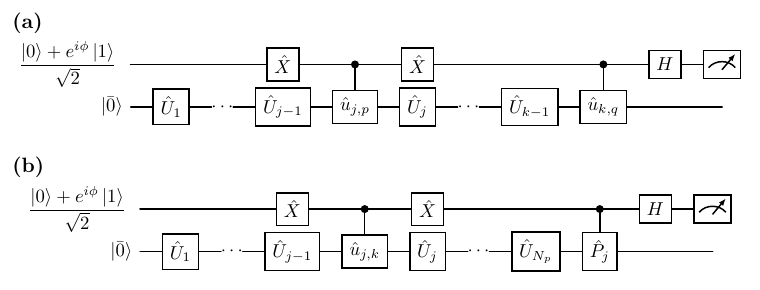}
    \caption{Quantum circuits for calculating (a) $\Re(e^{i\phi}\bra*{\bar{0}}\mathcal{\hat{U}}_{j,p}^\dag \mathcal{\hat{U}}_{k,q}\ket*{\bar{0}})$ and (b) $\Re(e^{i\phi}\bra*{\bar{0}} \mathcal{\hat{U}}_{j,k}^\dag \hat{P}_j \hat{U}\ket*{\bar{0}})$, which are required to obtain $M_{j,k}$ in Eq.~(\ref{eq:M}) and $C_{j}$ in Eq.~(\ref{eq:V}). The upper (lower) horizontal line represents the ancillary qubit (the system qubits). The initial state is prepared as $(\ket*{0}+e^{i \phi}\ket*{1})/\sqrt{2}$. $\hat X$ and $H$ denote the Pauli-X gate and the Hadamard gate, respectively, and $\hat{U}_j$ ($j=1,\cdots, N_{\mathrm{ p}}$) is a parametrized unitary gate constituting the variational circuit. The expectation values of the $Z$-measurement on the ancillary qubit yield (a) $\Re(e^{i\phi}\bra*{\bar{0}}\mathcal{\hat{U}}_{j,p}^\dag \mathcal{\hat{U}}_{k,q}\ket*{\bar{0}})$ and (b) $\Re(e^{i\phi}\bra*{\bar{0}} \mathcal{\hat{U}}_{j,k}^\dag \hat{P}_j \hat{U}\ket*{\bar{0}})$. }
    \label{fig:vite circuit}
\end{figure}

We now consider the derivative of each parameterized quantum gate. We assume that the derivative of $\hat{U}_j$ can be decomposed as
\begin{align}
    \frac{\partial \hat{U}_j}{\partial \theta_{j}} = \sum_{k} a_{j,k} \hat{U}_{j} \hat{u}_{j,k},
    \label{eq:expansion}
\end{align}
where $\hat{u}_{j,k}$ is a unitary operator and $a_{j,k}$ is a complex coefficient. 
\begin{align}
    \frac{\partial \ket*{\varphi(\bm{\theta}(\beta))}}{\partial \theta_{j}} = \sum_{k} a_{j,k} \mathcal{\hat{U}}_{j,k} \ket*{\bar{0}},
    \label{Eq:partialstate}
\end{align}
where
\begin{align}
    \mathcal{\hat{U}}_{j,k}= \hat{U}_{N_{\mathrm{ p}}}  \cdots \hat{U}_{j+1} \hat{U}_{j} \hat{u}_{j,k} \hat{U}_{j-1} \cdots  \hat{U}_{1}.
    \label{Eq: uki}
\end{align}
From Eq.~(\ref{Eq: uki}), the matrix element $M_{j,k}$ becomes
\begin{align}
    M_{j,k} = \sum_{p,q}\Re \left(a^{*}_{j,p}a_{k,q} \bra*{\bar{0}} \mathcal{\hat{U}}^\dagger_{j,p} \mathcal{\hat{U}}_{k,q} \ket*{\bar{0}}\right),
    \label{eq:M}
\end{align}
and the vector $\bm{C}$ is given by 
\begin{align}
    C_j = -\sum_{\mu,p} \Re \left(a_{j,p} h_\mu \bra*{\bar{0}} \hat{U}^\dagger (\bm{\theta})\hat{P}_\mu \hat{\mathcal U}_{j,p}\ket*{\bar{0}}\right),
    \label{eq:V}
\end{align}
where the Hamiltonian has been decomposed as $\hat H=\sum_\mu h_\mu \hat P_\mu$ with real coefficients $h_\mu$
and Pauli strings $\hat P_\mu$.
Each element in $M_{j,k}$ ($C_j$) can be efficiently calculated by quantum circuits shown in Fig.~\ref{fig:vite circuit} on the NISQ devices.

Since $M$ is a Hermitian ($N_{\mathrm{p}}\times N_{\mathrm{p}}$) matrix and $\bm{C}$ is an $N_{\mathrm{p}}$-dimensional vector, the number of independent components scales as $\mathcal{O}(N_{\mathrm{p}}^2)$. For $\bm{C}$, additional circuits are required due to the Hamiltonian decomposition.

%%%%%%%%%%%%%%%%%%%%%%%%%%%%%%%%%%%%%%%%%%%%%%%%%%%%%%%%%%%
%%%%%%%%%%%%%%%%%%%%%%%%%%%%%%%%%%%%%%%%%%%%%%%%%%%%%%%%%%%
\section{Overview of Variational Quantum-Neural Hybrid Eigensolver}\label{sec:VQNHE}
%%%%%%%%%%%%%%%%%%%%%%%%%%%%%%%%%%%%%%%%%%%%%%%%%%%%%%%%%%%
%%%%%%%%%%%%%%%%%%%%%%%%%%%%%%%%%%%%%%%%%%%%%%%%%%%%%%%%%%%
Before introducing our hybrid ITE algorithm in Sect.~\ref{sec:VQNHITE}, we briefly review VQNHE \cite{zhang2022variational}. This method, which combines a parameterized quantum circuit with a neural-network-generated non-unitary operator, forms the basis for enhancing the expressive power of variational algorithms for ground-state energy estimation. 
A key element of VQNHE is the introduction of a non-unitary operator
\begin{align}
    \hat{f}(\bm{\phi}) = \sum_{\mathclap{s \in \{0,1\}^{N_{\mathrm{q}}}}} f_{\bm{\phi}}(s)\ketbra{s},
\end{align}
where $f_{\bm{\phi}}(s)$ is a function computed by a NN, $\bm{\phi}$ denotes the NN parameters, and $s$ is a bitstring. For simplicity, $ f_{\bm{\phi}}(s) $ is assumed to be real-valued throughout this work, although complex-valued functions can also be defined.

The NN consists of an input layer, hidden layers, and an output layer. We assume that the NN has $N_{\mathrm{layer}}$ layers, and the activation vector in the $(j+1)$th layer is computed as  
\begin{equation}
    \bm{x}_{j+1} = g(W_j \bm{x}_j + \bm{b}_j) \ (j=1,\cdots, N_{\mathrm{layer}}-1),
\end{equation}
where $g$ is a nonlinear activation function, and $W_j$ and $\bm{b}_j$ denote the weight matrix and bias vector, respectively. In this formulation, the NN parameters are collectively denoted by $\bm{\phi}=\{W_j, \bm{b}_j\}$. 
After preparing a parameterized state $\ket*{\varphi (\bm{\theta})}= \hat{U}(\bm{\theta})\ket*{\bar{0}}$, we evaluate the expectation value of $\hat{H}=\sum_{j} h_j \hat{P}_j$ as
\begin{align}
    \expval*{\hat{H}}=\sum_j h_j \mathcal{N}(\bm{\lambda})^2\bra*{\tilde{\varphi}(\bm{\lambda})}\hat{P}_j\ket*{\tilde{\varphi}(\bm{\lambda})},
    \label{eq: expectation value with the neural network}
\end{align}
where $\bm{\lambda}= (\bm{\theta}, \bm{\phi})$, $\ket*{\tilde{\varphi}(\bm{\lambda})} = \hat{f}(\bm{\phi}) \ket*{\varphi(\bm{\theta})}$, and $\mathcal{N}(\bm{\lambda})=1/\sqrt{\braket{\tilde{\varphi}(\bm{\lambda})}{\tilde{\varphi}(\bm{\lambda})}}$.

When $\hat{P}_j$ consists only of the Pauli-$\hat{Z}$ operators, the numerator and denominator of Eq.~\eqref{eq: expectation value with the neural network} can be efficiently obtained by measuring all qubits in the computational basis and processing the resulting bitstrings $s$ with the NN. 

Next, we consider the case in which $\hat{P}_j$ contains the Pauli-$\hat{X}$ or Pauli-$\hat{Y}$ operators.
For clarity, we relabel one of the qubits acted on by the non-Pauli-$\hat{Z}$ operator as the zeroth qubit.
The operator $\hat{P}_j$ acts on a computational-basis state $\ket*{s}$ as
\begin{align}
    \hat{P}_j \ket*{s} = S(\tilde{s}) \ket*{\tilde{s}},
    \label{eq:example of transformed bitstring}
\end{align}
where $S(\tilde{s})$ is a phase factor $( \pm 1 )$ or $( \pm i )$, and $\tilde{s}$ represents the bitstring obtained from $s$ by the action of $\hat{P}_j$. 
For notational simplicity, we write the transformed bitstring and phase factor associated with $\hat{P}_j$ as $\tilde s$ and $S(\tilde s)$, respectively, omitting the subscript $j$.
For example, when $\hat{P}_j = \hat{X}_0 \hat{Y}_1 \hat{Z}_2$ and $s = 011 $, the transformed bitstring becomes $\tilde{s} = 101$.
Using Eq.~\eqref{eq:example of transformed bitstring}, $\hat{P}_j$ can be expressed as
\begin{align}
    \hat{P}_j=
    \sum_{\mathclap{\substack{
        s_0=0, \\
        s_{1:N_{\mathrm{q}}-1} \in \{0,1\}^{N_{\mathrm{q}}-1}
    }}}
    \Big(S(s)\ket*{s}\bra*{\tilde{s}} + S(\tilde{s})\ket*{\tilde{s}}\bra*{s} \Big),
    \label{eq: expression of Oi}
\end{align}
where the bitstring is decomposed as $s=(s_0,s_{1:N_{\mathrm{q}}-1})$.

The eigenstates of $\hat{P}_j$ corresponding to the eigenvalues $\pm 1$ are then given by
\begin{align}
    \ket*{+,s_{1:N_{\mathrm{q}}-1}}=\frac{1}{\sqrt{2}}\Big[S(0s_{1:N_{\mathrm{q}}-1})\ket*{0s_{1:N_{\mathrm{q}}-1}}+\ket*{1\tilde{s}_{1:N_{\mathrm{q}}-1}}\Big],
\end{align}
\begin{align}
    \ket*{-,s_{1:N_{\mathrm{q}}-1}}=\frac{1}{\sqrt{2}}\Big[S(0s_{1:N_{\mathrm{q}}-1})\ket*{0s_{1:N_{\mathrm{q}}-1}}-\ket*{1\tilde{s}_{1:N_{\mathrm{q}}-1}}\Big].
\end{align}
Here, we define $\ket*{0s_{1:N_{\mathrm{q}}-1}}=\ket*{s_0=0} \ket*{s_{1:N_{\mathrm{q}}-1}} $ and $\ket*{1s_{1:N_{\mathrm{q}}-1}}=\ket*{s_0=1}\ket*{s_{1:N_{\mathrm{q}}-1}}$.
Using the eigenstates $\ket*{\pm,s_{1:N_{\mathrm{q}}-1}}$ defined above, the matrix element $\bra*{\tilde{\varphi}(\bm{\lambda})}\hat{P}_j\ket*{\tilde{\varphi}(\bm{\lambda})}$ can be evaluated as
\begin{align}
    \bra*{\tilde{\varphi}(\bm{\lambda})}\hat{P}_j\ket*{\tilde{\varphi}(\bm{\lambda})}
    = \sum_{\mathclap{\substack{
        s_0=0,\\
        s_{1:N_{\mathrm{q}}-1}\in\{0,1\}^{N_{\mathrm{q}}-1}
    }}} (|\varphi_{+,s}|^2 - |\varphi_{-,s}|^2)
    f_{\bm{\phi}}(s) f_{\bm{\phi}}(\tilde{s}),
    \label{eq: computation of a non-diagonal term with NN}
\end{align}
where $\varphi_{\pm, s} = \langle \pm, s_{1:N_{\mathrm{q}}-1} | \varphi(\bm{\theta}) \rangle$.
In this approach, the ground-state energy is estimated by optimizing the parameters $\bm{\theta}$ and $\bm{\phi}$ using a gradient-based method. 

We can estimate $|\varphi_{\pm,s}|^2$ using a measurement circuit $V$ designed for the operator $\hat{P}_j$ after applying $\hat{U}(\bm{\theta})$. The circuit $V^{\dag}$ transforms $\ket*{0s_{1:N_{\mathrm{q}}-1}}$ and $\ket*{1s_{1:N_{\mathrm{q}}-1}}$ into $\ket*{+,s_{1:N_{\mathrm{q}}-1}}$ and $\ket*{-,s_{1:N_{\mathrm{q}}-1}}$, respectively, up to a global phase.
For example, when $\hat{P}_j = \hat{X}_0 \hat{Y}_1 \hat{Z}_2$, the measurement circuit is given by $\hat H_0\mathrm{C}\hat{Y}_{0,1}\mathrm{C}\hat{Z}_{0,2}$, followed by measurements in the computational basis, where $H_0$ denotes the Hadamard gate and $\mathrm{C}\hat{Y}_{0,1}$ denotes the controlled-$\hat{Y}$ gate, and $\mathrm{C}\hat Z_{0,2}$ denotes the controlled-$\hat Z$ gate.
In Sect.~\ref{sec:VQNHITE}, we extend the VQNHE framework to imaginary-time evolution. By deriving coupled variational update equations for both the circuit parameters $\bm{\theta}$ and the NN parameters $\bm{\phi}$, we construct a hybrid ITE scheme that achieves higher expressivity than conventional VITE.

%%%%%%%%%%%%%%%%%%%%%%%%%%%%%%%%%%%%%%%%%%%%%%%%%%%%%%%%%%%
%%%%%%%%%%%%%%%%%%%%%%%%%%%%%%%%%%%%%%%%%%%%%%%%%%%%%%%%%%%
\section{Variational Quantum-Neural Hybrid Imaginary-Time Evolution}\label{sec:VQNHITE}
%%%%%%%%%%%%%%%%%%%%%%%%%%%%%%%%%%%%%%%%%%%%%%%%%%%%%%%%%%%
%%%%%%%%%%%%%%%%%%%%%%%%%%%%%%%%%%%%%%%%%%%%%%%%%%%%%%%%%%%

In this section, we present our proposed hybrid imaginary-time evolution algorithm, VQNHITE, which integrates the variational quantum circuit of VITE with a neural-network-based non-unitary operator. The procedure consists of two stages: (i) preparing suitable initial parameters $\bm{\theta}$ and $\bm{\phi}$ at a small imaginary-time step $\beta=\delta \beta$, and (ii) performing imaginary-time evolution for general $\beta$ using a joint update of $\bm{\theta}$ and $\bm{\phi}$. Sect.~\ref{sec:init} describes the initialization step, and Sect.~\ref{sec:update} provides the hybrid update equations.

%%%%%%%%%%%%%%%%%%%%%%%%%%%%%%%%%%%%%%%%%%%%%%%%%%%%%%%%%%%
\subsection{Preparation of Parameters}\label{sec:init}
%%%%%%%%%%%%%%%%%%%%%%%%%%%%%%%%%%%%%%%%%%%%%%%%%%%%%%%%%%%
At $\beta=0$, we choose $\bm{\theta}=\bm{0}$ so that the parameterized unitary $\hat U(\bm{\theta})$ reduces to the identity operator.
For the NN, we initialize the first-layer weight matrix as $W_1=0$, while the biases and the remaining NN parameters are initialized randomly. This makes the NN output independent of the input bitstring $s$, and hence the NN-based operator is proportional to the identity operator. The proportionality factor is removed by the normalization of $\ket{\Phi(\bm{\lambda})}$. Therefore, the normalized hybrid state at $\beta=0$ reduces to the initial variational state.

If we directly apply the update rule in Eq.~\eqref{eq:renew_param} to such initial parameters, the randomness in $W_k$ and $\bm{b}_k$ often leads to degraded fidelity during the imaginary-time evolution.
To mitigate this issue, we optimize $\bm{\theta}$ and $\bm{\phi}$ for $\beta=\delta\beta$ by using the lower-order short-time approximation in Eq.~\eqref{eq:short_time_approx}.
For sufficiently small $\delta\beta$, this approximation is given by
\begin{align}
    e^{-\hat{H}\delta\beta}\simeq I-\hat{H}\delta\beta+\frac{1}{2}(\hat{H}\delta\beta)^2=\sum_{\mu} e_{\mu}\hat{P}_{\mu},
    \label{eq:short_time_approx}
\end{align}
where $\hat{P}_{\mu}$ denotes a Pauli string and $e_{\mu}$ is its
coefficient. This approximation is employed only for the initialization step at $\beta=\delta\beta$.
% For $\beta>\delta\beta$, the parameters are updated according to the variational equations introduced in the next subsection.

We introduce the cost function
\begin{align}
    F_{\mathrm{cost}}=1-\abs{\braket{\psi(\delta\beta)}{\Phi(\bm{\lambda})}}^2,
\end{align}
where $\ket*{\Phi(\bm{\lambda})}=\mathcal N(\bm{\lambda})\ket*{\tilde{\varphi}(\bm{\lambda})}$. 
The target state $\ket{\psi(\delta\beta)}$ is obtained by applying Eq.~\eqref{eq:short_time_approx} to $\ket{\psi(0)}$ and normalizing the resulting state.

For notational convenience, we define $S(\bm{\lambda}):=\braket{\psi(\delta\beta)}{\tilde{\varphi}(\bm{\lambda})}$, 
$G_{\lambda_a}(\bm{\lambda}):=\bra*{\psi(\delta\beta)}\partial_{\lambda_a}\ket*{\tilde{\varphi}(\bm{\lambda})}$, and $D_{\lambda_a}(\bm{\lambda}):=\Re\!\left[\braket{\tilde{\varphi}(\bm{\lambda})}{\partial_{\lambda_a}\tilde{\varphi}(\bm{\lambda})}\right]$, 
where $\bm{\lambda}=(\bm{\theta},\bm{\phi})$ is the combined parameter
vector and $\lambda_a$ denotes its $a$th component.
Then
\begin{align}
    \partial_{\lambda_a}\mathcal N(\bm{\lambda})
    =
    -\mathcal N(\bm{\lambda})^3 D_{\lambda_a}(\bm{\lambda}),
\end{align}
and the gradient of the cost function is written as
\begin{align}
    \partial_{\lambda_a}F_{\mathrm{cost}}
    &=
    2\mathcal N(\bm{\lambda})^4
    D_{\lambda_a}(\bm{\lambda})\,|S(\bm{\lambda})|^2
    \notag\\
    &\quad-
    2\mathcal N(\bm{\lambda})^2
    \Re\!\left[
    S(\bm{\lambda})^*\,G_{\lambda_a}(\bm{\lambda})
    \right].
    \label{eq:dFcost_lambda}
\end{align}
For the NN and circuit parameters, we have
\begin{align}
    D_{\phi_j}
    &=
    \Re\left(
    \sum_{s \in \{0,1\}^{N_q}}
    f_{\bm{\phi}}(s)\,\partial_{\phi_j}f_{\bm{\phi}}(s)\,
    |\langle \varphi(\bm{\theta}) | s \rangle|^2
    \right),
    \label{eq:D_phi}\\
    D_{\theta_j}
    &=
    \Re\left(
    \sum_{s \in \{0,1\}^{N_q}}
    f_{\bm{\phi}}(s)^2
    \langle \bar{0} | \partial_{\theta_j}\hat{U}^{\dagger} | s \rangle
    \langle s | \varphi(\bm{\theta}) \rangle
    \right).
    \label{eq:D_theta}
\end{align}
The explicit expressions for
$S(\bm{\lambda})$, $G_{\lambda_a}(\bm{\lambda})$, and $D_{\lambda_a}(\bm{\lambda})$ are given in Appendix~\ref{sec: Quantum circuits for evaluating the derivatives}.

We update all parameters using gradient descent $\lambda_a \leftarrow \lambda_a - \eta\,\partial_{\lambda_a}F_{\mathrm{cost}}$, with learning rate $\eta$.
After a fixed number of iterations, the resulting optimized parameters are taken as $\bm{\lambda}(\beta=\delta\beta)$, namely $\bm{\theta}(\beta=\delta\beta)$ and $\bm{\phi}(\beta=\delta\beta)$, which serve as the initialization for the subsequent hybrid imaginary-time evolution.

%%%%%%%%%%%%%%%%%%%%%%%%%%%%%%%%%%%%%%%%%%%%%%%%%%%%%%%%%%%
\subsection{Updating Parameters}\label{sec:update}
%%%%%%%%%%%%%%%%%%%%%%%%%%%%%%%%%%%%%%%%%%%%%%%%%%%%%%%%%%%
We now describe the parameter update for general $\beta>\delta\beta$. 
Extending Eq.~\eqref{eq:MC} from the circuit parameter vector
$\bm{\theta}$ to the combined parameter vector $\bm{\lambda}$, the generalized variational update equation is written as
\begin{align}
    \sum_b \mathcal M_{\lambda_a\lambda_b} \dot{\lambda}_b
    =
    \mathcal C_{\lambda_a},
    \label{eq:lambda_update}
\end{align}
where $\mathcal M_{\lambda_a\lambda_b}:=\Re\!\left[\braket{\partial_{\lambda_a}\Phi(\bm{\lambda})}{\partial_{\lambda_b}\Phi(\bm{\lambda})}\right],$ and $ \mathcal C_{\lambda_a}:=-\Re\!\left[\bra*{\partial_{\lambda_a}\Phi(\bm{\lambda})}\hat{H}\ket*{\Phi(\bm{\lambda})}\right]$.
Here the indices $a$ and $b$ run over all components of the combined parameter vector $\bm{\lambda}$; thus $\lambda_a$ denotes the $a$-th component of $\bm{\lambda}$.

Using the normalized hybrid state $\ket*{\Phi(\bm{\lambda})}$ and the quantity $D_{\lambda_a}(\bm{\lambda})$,
we define $K_{\lambda_a\lambda_b}(\bm{\lambda})
:=\braket{\partial_{\lambda_a}\tilde{\varphi}(\bm{\lambda})}
{\partial_{\lambda_b}\tilde{\varphi}(\bm{\lambda})}$
and
$L_{\lambda_a}(\bm{\lambda})
:=\bra*{\partial_{\lambda_a}\tilde{\varphi}(\bm{\lambda})}
\hat{H}\ket*{\tilde{\varphi}(\bm{\lambda})}$, whose matrix and vector elements are written as
\begin{align}
    \mathcal M_{\lambda_a\lambda_b}
    &=
    \mathcal N(\bm{\lambda})^2
    \Re\!\left[K_{\lambda_a\lambda_b}(\bm{\lambda})\right]
    -
    \mathcal N(\bm{\lambda})^4
    D_{\lambda_a}(\bm{\lambda})D_{\lambda_b}(\bm{\lambda}),
    \label{eq:lambda_metric}
\end{align}
and
\begin{align}
    \mathcal C_{\lambda_a}
    &=
    -\mathcal N(\bm{\lambda})^2
    \Re\!\left[L_{\lambda_a}(\bm{\lambda})\right]
    +
    \mathcal N(\bm{\lambda})^2
    D_{\lambda_a}(\bm{\lambda})\,E(\bm{\lambda}),
    \label{eq:lambda_force}
\end{align}
where $E(\bm{\lambda}):=\expval{\hat{H}}{\Phi(\bm{\lambda})}$.

The update at each imaginary-time step is then given by
\begin{align}
    \bm{\lambda}(\beta+\delta\beta)
    \simeq
    \bm{\lambda}(\beta)
    +
    \mathcal M(\beta)^{-1}\mathcal C(\beta)\,\delta\beta.
\end{align}
The explicit expressions for
$K_{\lambda_a\lambda_b}(\bm{\lambda})$ and
$L_{\lambda_a}(\bm{\lambda})$
are given in Appendix~\ref{sec: Quantum circuits for evaluating the derivatives}.

The components required to construct $\mathcal M_{\lambda_a\lambda_b}$ and $\mathcal C_{\lambda_a}$ can be evaluated using the Hadamard test, as detailed in Appendix~\ref{sec: Quantum circuits for evaluating the derivatives}. Although this hybrid extension requires a larger variety of quantum circuits to estimate $\mathcal M_{\lambda_a\lambda_b}$ and $\mathcal C_{\lambda_a}$ than the conventional VITE, overall sampling costs remain manageable. As shown in Sect.~\ref{sec:result}, this approach yields a substantially higher fidelity to the exact ITE state. Following the same strategy as VQNHE, our method can also compute expectation values with respect to the state virtually produced by our method.

%%%%%%%%%%%%%%%%%%%%%%%%%%%%%%%%%%%%%%%%%%%%%%%%%%%%%%%%%%%
%%%%%%%%%%%%%%%%%%%%%%%%%%%%%%%%%%%%%%%%%%%%%%%%%%%%%%%%%%%
\section{Numerical Results}\label{sec:result}
%%%%%%%%%%%%%%%%%%%%%%%%%%%%%%%%%%%%%%%%%%%%%%%%%%%%%%%%%%%
%%%%%%%%%%%%%%%%%%%%%%%%%%%%%%%%%%%%%%%%%%%%%%%%%%%%%%%%%%%
\begin{figure*}[t]
    \centering
    \includegraphics[width=\textwidth]{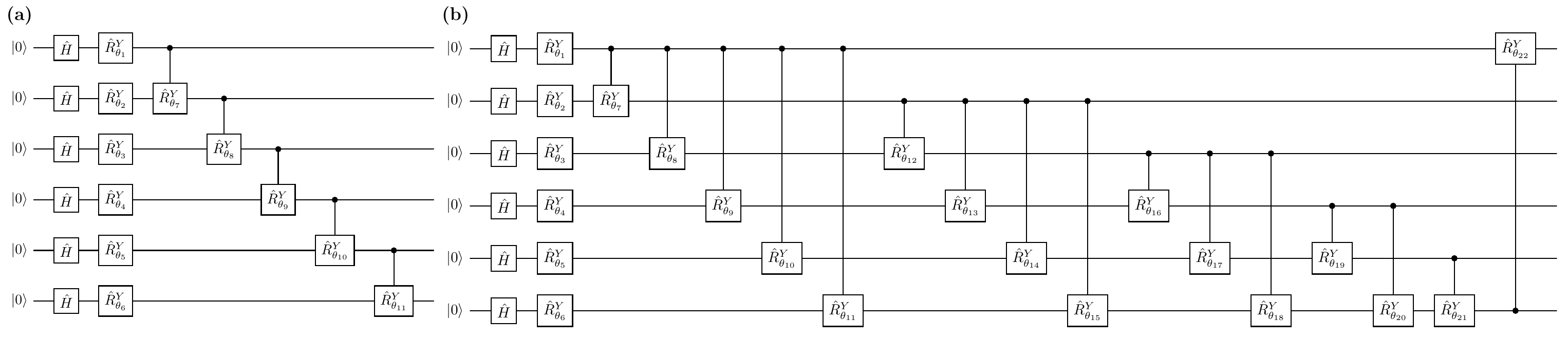}
    \caption{The ansatz circuit for $N=6$ qubits. Panels (a) and (b) show circuits containing only single-qubit rotation gates together with (a) nearest-neighbor interactions or (b) all-to-all interactions. $R^Y_{\theta_i}$ denotes a rotation along the $Y$-axis, and $\theta_i$ represents the parameter of the $i$th rotation angle.}
    \label{fig:ansatz}
\end{figure*}

In this section, we demonstrate that our VQNHITE method achieves higher accuracy than conventional VITE. As a benchmark, we consider the fidelity between the variational state produced by either VQNHITE or VITE and the ideal imaginary-time-evolved state. 
Although our method does not physically prepare the normalized hybrid state $\ket*{\Phi}$, it enables us to evaluate observables as if the state were prepared.
However, since the fidelity serves as a strict metric, we evaluate it through numerical simulations. The initial state is taken as the all-plus state $\ket*{+\cdots+}$.

We consider the Heisenberg Hamiltonian with a longitudinal field under open boundary conditions:
\begin{align} 
    \notag&\hat{H}=\hat{H}_{\mathrm{Heisenberg}}+\hat{H}_{\mathrm{field}}\\
    &=J \sum_{j=0}^{N-2} ( \hat{\sigma}_j^x \hat{\sigma}_{j+1}^x + \hat{\sigma}_j^y \hat{\sigma}_{j+1}^y + \hat{\sigma}_j^z \hat{\sigma}_{j+1}^z )+\sum_{j=0}^{N-1} h_j\hat{\sigma}_j^z, 
    \label{eq:heisenberg_model} 
\end{align} 
where $N$ denotes the number of qubits, $J$ represents the strength of the exchange interaction, and $h_j$ indicates the strength of the longitudinal magnetic field acting on the $j$th qubit. In our numerical calculations, we set $N=6$ and $8$, $J=-1$, and assigned the values of $h_j\sim\mathrm{Unif}[-1,1]$ randomly. 

For the gradient descent used during the optimization of the initial parameters, we performed $50$ iterations with a learning rate of $0.1$.
The NN model we utilized is a fully connected NN with two hidden layers of width $N$ and $N/2$ equipped with the $\tanh$ activation function.

In actual devices, interconnectivity is limited. Thus, we consider two types of quantum circuits: the circuit in Fig.~\ref{fig:ansatz}(a), which includes only nearest-neighbor interactions, and the circuit in Fig.~\ref{fig:ansatz}(b), which consists of all-to-all interactions.

\begin{figure}[!tbp]
    \centering
    \includegraphics[width=0.85\columnwidth]{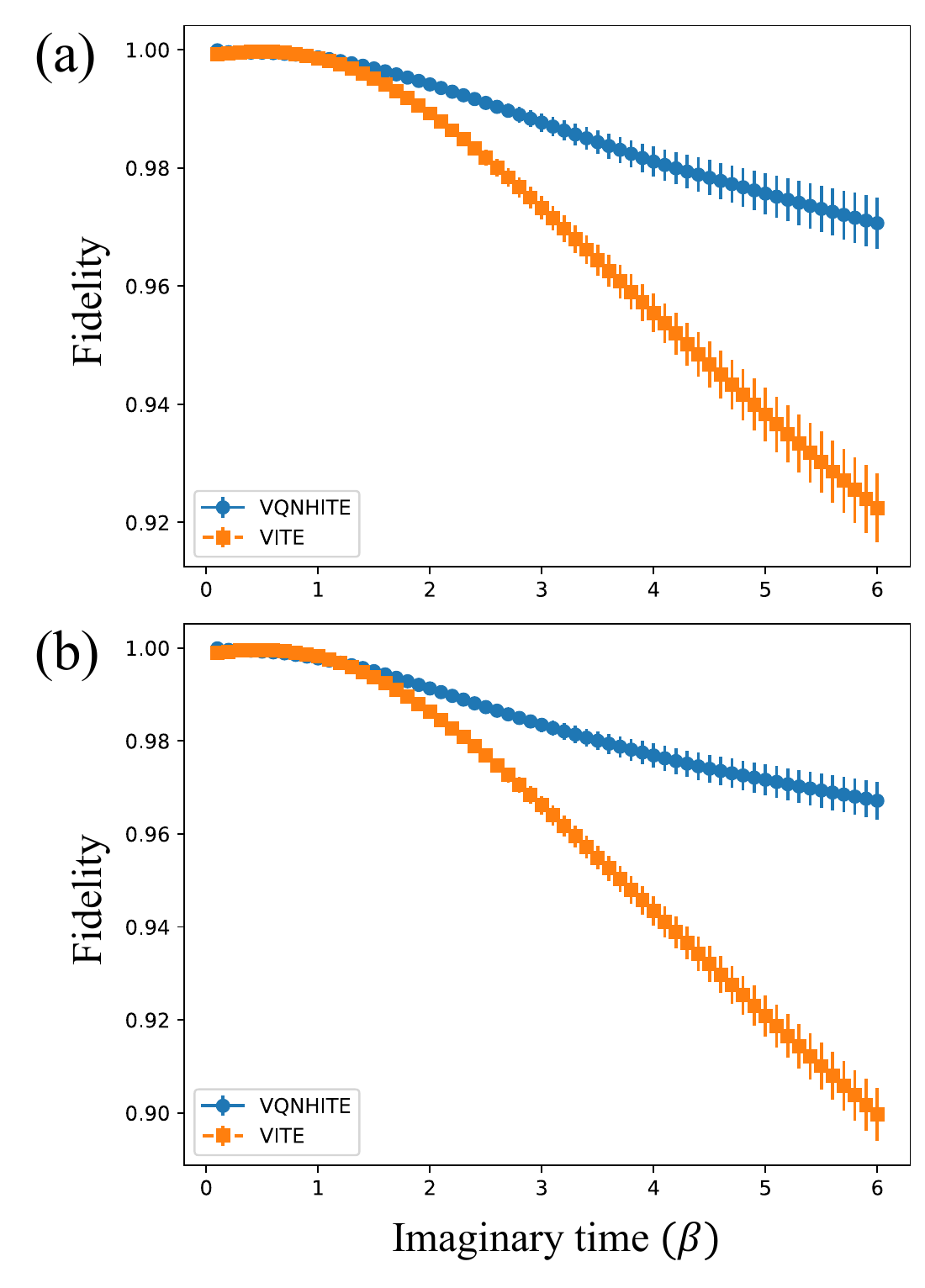}
    \caption{(Color online) Fidelity of the variational states obtained by our VQNHITE (circular markers) and by VITE (square markers) with respect to the exact imaginary-time-evolved state $\ket{\psi(\beta)}$, plotted as a function of the imaginary time $\beta$. The horizontal axis spans $\beta\in[0.1,6]$ with step size $\Delta\beta=0.1$. Results are shown for the nearest-neighbor ansatz in Fig.~\ref{fig:ansatz}(a). Panels (a) and (b) correspond to $N=6$ and $N=8$, respectively. We set $J=-1$, and sample the field strength as $h_j\sim\mathrm{Unif}[-1,1]$. Each data point represents the sample mean over $100$ independent runs, and the error bars denote the standard error of the mean. In each run, an independent realization of $\{h_j\}$ is generated, and both VITE and VQNHITE are evaluated for the same Hamiltonian realization. The reference state $|\psi(\beta)\rangle$ is computed classically by normalizing $\exp(-\beta \hat H)|\psi(0)\rangle$. The fidelities for VITE and VQNHITE are defined as $F(\beta)=|\langle\psi(\beta)\vert\varphi(\bm{\theta})\rangle|^2$ and $F(\beta)=|\langle\psi(\beta)\vert\Phi(\bm{\lambda})\rangle|^2$, respectively.}
    \label{fig:fidelity_nn}
\end{figure}

\begin{figure}[!tbp]
    \centering
    \includegraphics[width=0.85\columnwidth]{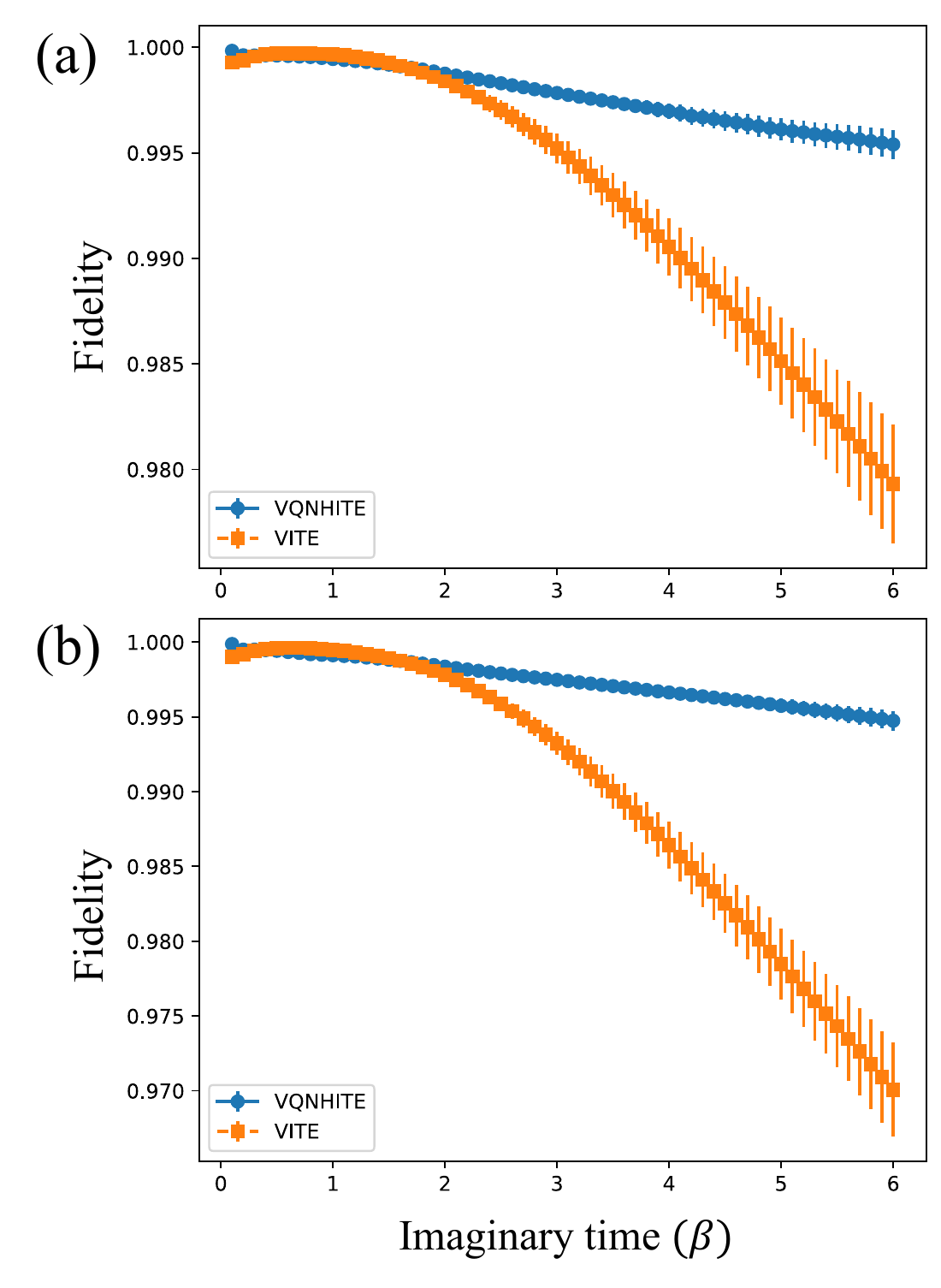}
    \caption{(Color online) Fidelity of the variational states obtained by our VQNHITE (circular markers) and by VITE (square markers) with respect to the exact imaginary-time-evolved state $\ket{\psi(\beta)}$, plotted as a function of the imaginary time $\beta$. The horizontal axis spans $\beta\in[0.1,6]$ with step size $\Delta\beta=0.1$. Results are shown for the all-to-all ansatz in Fig.~\ref{fig:ansatz}(b). Panels (a) and (b) correspond to $N=6$ and $N=8$, respectively. We set $J=-1$, and sample the field strength as $h_j\sim\mathrm{Unif}[-1,1]$. Each data point represents the sample mean over $100$ independent runs, and the error bars denote the standard error of the mean. In each run, an independent realization of $\{h_j\}$ is generated, and both VITE and VQNHITE are evaluated for the same Hamiltonian realization. The reference state $\ket{\psi(\beta)}$ is computed classically by normalizing $\exp(-\beta \hat H)\ket{\psi(0)}$. The fidelities for VITE and VQNHITE are defined as $F(\beta)=|\langle\psi(\beta)\vert\varphi(\bm{\theta})\rangle|^2$ and $F(\beta)=|\langle\psi(\beta)\vert\Phi(\bm{\lambda})\rangle|^2$, respectively.}
    \label{fig:fidelity_all}
\end{figure}

Figures~\ref{fig:fidelity_nn} and~\ref{fig:fidelity_all} plot the fidelity on the vertical axis and the imaginary time on the horizontal axis for the cases of nearest-neighbor and all-to-all interactions, respectively. 
The imaginary time is incremented in steps of $0.1$, and the plot shows results from $0.1$ to $6$. 
The fidelity calculated using the VITE, defined as $|\langle\psi(\beta)\vert\varphi(\bm{\theta})\rangle|^2$, is represented by square markers, while that obtained using our VQNHITE, defined as $|\langle\psi(\beta)\vert \Phi(\bm{\lambda})\rangle|^2$, is represented by circular markers. 
Here, the state $\ket*{\psi(\beta)}$ denotes the ideal imaginary-time-evolved state used only for benchmarking, and it is computed classically by exact state-vector simulation as the normalized state proportional to $\exp(-\beta \hat{H})\ket*{\psi(0)}$.
In a realistic implementation, the target state $\ket*{\psi(\delta\beta)}$ used in the initialization step can be estimated using the lower-order short-time approximation in Eq.~\eqref{eq:short_time_approx}.
% Although the initial state $\ket*{\psi(\delta \beta)}$ can be estimated using a lower-order short-time approximation in a realistic implementation as described in Appendix~\ref{sec: Quantum circuits for evaluating the derivatives}
In the present numerical simulations, we instead use the exact target state at $\beta=\delta\beta$ for simplicity. 

For each data point, we performed 100 independent runs. In each run, the longitudinal fields $\{h_j\}$ in Eq.~\eqref{eq:heisenberg_model} were independently sampled from $\mathrm{Unif}[-1,1]$, and both VITE and VQNHITE were evaluated for the same Hamiltonian realization. The plotted values show the sample mean over the $100$ runs, and the error bars denote the standard error of the mean. For VQNHITE, the run-to-run variability also includes the effect of the random NN initialization used in the implementation.

Figures~\ref{fig:fidelity_nn} and \ref{fig:fidelity_all} show that our VQNHITE consistently achieves higher fidelity than VITE, and that the performance gap increases with $\beta$.
This trend is observed for both the nearest-neighbor and all-to-all ansatz circuits, even though both methods use the same circuit depth for the parameterized quantum circuit. 
The gradual decrease of the fidelity with increasing $\beta$ is due to the limited expressive power of the fixed-depth variational ansatz and the accumulation of approximation errors in the McLachlan variational principle.
In the present simulations, the initialization at $\beta=\delta\beta$ is also performed using the exact target state rather than the lower-order short-time approximation in Eq.~\eqref{eq:short_time_approx}.
Therefore, the observed fidelity degradation at larger $\beta$ should not be attributed to the short-time approximation itself.
The discrepancy between VITE and VQNHITE is mainly attributed to the additional expressive power provided by the NN-assisted non-unitary layer in VQNHITE. The comparison between the nearest-neighbor and all-to-all ansatz circuits further indicates that circuit connectivity affects the severity of fidelity decay.
When the nearest-neighbor ansatz is replaced with the all-to-all one, the fidelity decay becomes weaker in both the VITE and VQNHITE methods.
At the same time, VQNHITE consistently maintains higher fidelity during the dynamics than VITE in both connectivity settings.
These observations suggest that the fidelity degradation is primarily caused by the limited expressive power of the variational ansatz. The NN-assisted non-unitary layer enhances the expressivity and thereby reduces the approximation error relative to standard VITE.
In other words, as $\beta$ increases, the exact imaginary-time-evolved state gradually moves into a region of Hilbert space where states are harder to represent with a fixed-depth ansatz. This limitation is partially alleviated in VQNHITE by the additional non-unitary flexibility provided by the NN.

%%%%%%%%%%%%%%%%%%%%%%%%%%%%%%%%%%%%%%%%%%%%%%%%%%%%%%%%%%%
%%%%%%%%%%%%%%%%%%%%%%%%%%%%%%%%%%%%%%%%%%%%%%%%%%%%%%%%%%%
\section{Conclusion}\label{sec:discussion}
%%%%%%%%%%%%%%%%%%%%%%%%%%%%%%%%%%%%%%%%%%%%%%%%%%%%%%%%%%%
%%%%%%%%%%%%%%%%%%%%%%%%%%%%%%%%%%%%%%%%%%%%%%%%%%%%%%%%%%%
We have proposed VQNHITE, which augments VITE with a neural-network-based non-unitary operation. By incorporating the NN, the proposed framework enhances the expressive power of the variational ansatz, thereby enabling a more accurate approximation of ITE under limited quantum resources. Numerical simulations confirm that VQNHITE consistently achieves higher fidelity with the exact imaginary-time-evolved states than conventional VITE across different circuit connectivities. 

% In this work, we introduced a lower-order short-time approximation in Eq.~\eqref{eq:short_time_approx} for the initialization step in realistic case, which provides a simple and scalable starting point and helps stabilize the subsequent imaginary-time updates. Although the proposed VQNHITE framework introduces additional computational time and sampling costs compared with conventional VITE, the overall overhead remains manageable for near-term devices~\cite{zhang2022variational}. 
In this work, we introduced an initial-parameter optimization at $\beta=\delta\beta$ to reduce the sensitivity to randomly initialized NN parameters. In a realistic implementation, the target state used in this initialization step can be constructed using the lower-order short-time approximation in Eq.~\eqref{eq:short_time_approx}. This approximation is used only for the initialization step, while the subsequent imaginary-time evolution is governed by the McLachlan update equations.
More generally, the choice of initialization is not unique, and developing alternative scalable initialization strategies that further improve the stability and performance of the method remains an important direction for future work.

Compared with conventional VITE, VQNHITE requires additional costs for the initial-parameter optimization at $\beta=\delta\beta$ and for evaluating $\mathcal{M}_{\lambda_a\lambda_b}$ and $\mathcal{C}_{\lambda_a}$ in the enlarged parameter space $\bm{\lambda}=(\bm{\theta},\bm{\phi})$. 
These required quantities can be evaluated by the measurement procedures described in Appendix~\ref{sec: Quantum circuits for evaluating the derivatives}. 

Our strategy can also evaluate expectation values of observables following the VQNHE measurement strategy, for which the overhead associated with non-unitary post-processing was shown to scale polynomially in Ref.~\cite{zhang2022variational}.
% The compact forms of $\mathcal{M}_{\lambda_a\lambda_b}$ and $\mathcal{C}_{\lambda_a}$ are given in Eqs.~\eqref{eq:lambda_metric} and \eqref{eq:lambda_force}, respectively, whereas their explicit component expressions and the corresponding evaluation circuits are presented in Appendix~\ref{sec: Quantum circuits for evaluating the derivatives}.

Future work includes extending VQNHITE to larger-scale systems and exploring alternative neural-network architectures that may further reduce sampling variance or improve generalization. We believe that the hybrid strategy developed here provides a promising direction for accurate imaginary-time simulation on NISQ hardware.

%%%%%%%%%%%%%%%%%%%%%%%%%%%%%%%%%%%%%%%%%%%%%%%%%%%%%%%%%%%%%%%%%%%%%%%%%%%%%%%%%%%%%%%%%%%%%%%
\begin{acknowledgments}
This work was supported by JST Moonshot R\&D Grant No. JPMJMS226C and JSPS KAKENHI Grant No. JP26KJ2001. H.K.’s research stay in Germany was supported by JST ASPIRE Grant No. JPMJAP24C2.
\end{acknowledgments}
%%%%%%%%%%%%%%%%%%%%%%%%%%%%%%%%%%%%%%%%%%%%%%%%%%%%%%%%%%%%%%%%%%%%%%%%%%%%%%%%%%%%%%%%%%%%%%%

\appendix
\setcounter{secnumdepth}{3}
\section{Quantum Circuits for Evaluating the Derivatives}
\label{sec: Quantum circuits for evaluating the derivatives}

\begin{figure*}[t]
\centering
\includegraphics[width=0.8\textwidth]{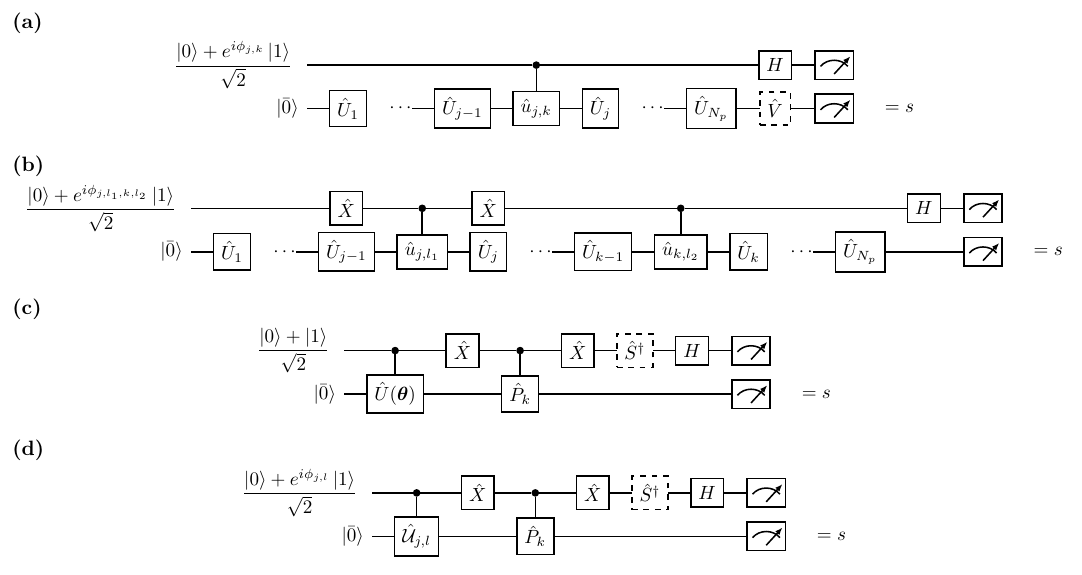}
\caption{
    Quantum circuits for calculating:
    (a)-1
    $e^{i\phi_{j,k}}\braket{\varphi(\bm{\theta})}{s}\bra*{s}\mathcal{\hat{U}}_{j,k}\ket*{\bar{0}}$,
    without the measurement circuit $V$;
    (a)-2
    $\Re\!\left(e^{i\phi_{j,l}}\varphi_{\pm,s}^{\theta_j,l}\right)$,
    with the measurement circuit $V$;
    (b)
    $\Re\!\left(
    e^{i\phi_{j,l_1,k,l_2}}
    \bra*{\bar{0}}\mathcal{\hat{U}}_{j,l_1}^{\dag}\ket*{s}
    \bra*{s}\mathcal{\hat{U}}_{k,l_2}\ket*{\bar{0}}
    \right)$;
    (c) $S(\bm{\lambda})$ and $G_{\phi_k}(\bm{\lambda})$;
    (d) $G_{\theta_j}(\bm{\lambda})$.
    To extract the real part in (c) and (d), the ancillary qubit is
    measured in the $\hat X$ basis using the Hadamard gate, since $\hat{H}\hat{Z}\hat{H} = \hat{X}$.
    To extract the imaginary part, the phase gate $\hat S^{\dagger}$ is
    inserted before the Hadamard gate so that the measurement corresponds to the $\hat Y$ basis, since $\hat S\hat{X}\hat S^{\dagger}=\hat Y$.
    The upper (lower) horizontal line represents the ancillary qubit
    (the system qubits).
    }
\label{Fig: circuit}
\end{figure*}

In this Appendix, we present quantum circuits for the quantities introduced in Sects.~\ref{sec:init} and~\ref{sec:update}. 
For the initialization step in Sect.~\ref{sec:init}, these include
the overlap
$S(\bm{\lambda})
=\braket{\psi(\delta\beta)}{\tilde{\varphi}(\bm{\lambda})}$,
its derivatives
$G_{\phi_j}(\bm{\lambda})$ and $G_{\theta_j}(\bm{\lambda})$,
and the normalization-related terms
$D_{\phi_j}(\bm{\lambda})$ and $D_{\theta_j}(\bm{\lambda})$.
We also summarize the matrix elements entering the update equations
in Sect.~\ref{sec:update}. 
Throughout this Appendix, we express complex coefficients in polar form $a_{j,k}= r_{j,k}e^{i\phi_{j,k}}$. This representation is convenient because the phase factor $e^{i\phi_{j,k}}$ can be directly incorporated into the Hadamard test, while the amplitude $r_{j,k}$ is treated as a classical weight. As a result, the extraction of real and imaginary parts using the circuits in Fig.~\ref{Fig: circuit} becomes straightforward.

\subsection{Quantities for the Initialization Step}
The quantity $D_{\phi_j}$ can be evaluated by measuring all qubits of
the state $\ket*{\varphi(\bm{\theta})}$ in the computational basis.
By inserting the obtained bitstring $s$ into $f_{\bm{\phi}}(s)\partial_{\phi_j} f_{\bm{\phi}}(s)$ and averaging over the measurement outcomes, we obtain $D_{\phi_j}$.

Using Eq.~\eqref{Eq:partialstate}, $D_{\theta_j}$ can be rewritten as
\begin{align}
    D_{\theta_j}=\Re\left(\sum_{s \in \{0,1\}^{N_q}} \sum_{k}a_{j,k}{|f_{\bm{\phi}}(s)|}^2\braket{\varphi(\bm{\theta})}{s}\bra*{s}\hat{\mathcal U}_{j,k}\ket*{\bar 0}\right),
    \label{eq:A1}
\end{align}
where $a_{j,k}=r_{j,k}e^{i\phi_{j,k}}$.
The real part of $e^{i\phi_{j,k}}\braket{\varphi(\bm{\theta})}{s}\bra*{s}\hat{\mathcal U}_{j,k}\ket*{\bar 0}$ is obtained using the circuit in Fig.~\ref{Fig: circuit}(a)-1, where the measurement circuit $V$ is not applied.

We next consider the overlap
\begin{align}
    S(\bm{\lambda})
    :=
    \braket{\psi(\delta\beta)}{\tilde{\varphi}(\bm{\lambda})}.
\end{align}
We define the target-state normalization factor as
\begin{equation}
    \mathcal N_{\mathrm{tar}}
    :=\left\|\sum_\mu e_\mu\hat P_\mu\ket*{\bar 0}\right\|^{-1}.
    \label{eq:target_normalization}
\end{equation}
Using the lower-order short-time approximation in Eq.~\eqref{eq:short_time_approx}, the overlap is written as
\begin{align}
    S(\bm{\lambda})
    =
    \mathcal N_{\mathrm{tar}}
    \sum_{\mu}
    \sum_{s\in\{0,1\}^{N_q}}
    e_{\mu}f_{\bm{\phi}}(s)\,
    \bra*{\bar 0}\hat{P}_{\mu}\ket*{s}
    \braket{s}{\varphi(\bm{\theta})}.
    \label{eq:target_overlap_expansion}
\end{align}
The real and imaginary parts of
$\bra*{\bar 0}\hat{P}_{\mu}\ket*{s}\braket{s}{\varphi(\bm{\theta})}$
are evaluated using the circuit in Fig.~\ref{Fig: circuit}(c).

Similarly, the derivative with respect to $\phi_j$ is written as
\begin{align}
    G_{\phi_j}(\bm{\lambda})
    &:=
    \bra*{\psi(\delta\beta)}
    \partial_{\phi_j}\ket*{\tilde{\varphi}(\bm{\lambda})}
    \notag\\
    &=
    \mathcal N_{\mathrm{tar}}
    \sum_{\mu}
    \sum_{s\in\{0,1\}^{N_q}}
    e_{\mu}\,
    \partial_{\phi_j}f_{\bm{\phi}}(s)\,
    \bra*{\bar 0}\hat{P}_{\mu}\ket*{s}
    \braket{s}{\varphi(\bm{\theta})},
\end{align}
whose real and imaginary parts are also evaluated with the circuit in
Fig.~\ref{Fig: circuit}(c).

Finally, the derivative with respect to $\theta_j$ is written as
\begin{align}
    G_{\theta_j}(\bm{\lambda})
    &:=
    \bra*{\psi(\delta\beta)}
    \partial_{\theta_j}\ket*{\tilde{\varphi}(\bm{\lambda})}
    \notag\\
    &=
    \mathcal N_{\mathrm{tar}}
    \sum_{\mu,l}
    \sum_{s\in\{0,1\}^{N_q}}
    e_{\mu}\,
    f_{\bm{\phi}}(s)\,
    a_{j,l}\,
    \bra*{\bar 0}\hat{P}_{\mu}\ket*{s}
    \bra*{s}\hat{\mathcal U}_{j,l}\ket*{\bar 0}.
    \label{eq:target_circuit_derivative}
\end{align}
Since both the real and imaginary parts of
$e^{i\phi_{j,l}}
\bra*{\bar 0}\hat{P}_{\mu}\ket*{s}
\bra*{s}\hat{\mathcal U}_{j,l}\ket*{\bar 0}$
can be evaluated using the circuit in Fig.~\ref{Fig: circuit}(d),
$G_{\theta_j}(\bm{\lambda})$ can also be computed with the same circuit.

\subsection{Matrix Elements for the Update Equations}

For the update equations in Sect.~\ref{sec:update}, we require
$K_{\lambda_a\lambda_b}(\bm{\lambda})$ and
$L_{\lambda_a}(\bm{\lambda})$. The following expressions give their components for $\lambda_a,\lambda_b \in \{\phi_j,\theta_j\}$.
Among these, $K_{\phi_j\phi_k}(\bm{\lambda})$ is given by
\begin{align}
    K_{\phi_j\phi_k}(\bm{\lambda})
    =
    \sum_{s\in\{0,1\}^{N_q}}
    \partial_{\phi_j}f_{\bm{\phi}}(s)\,
    \partial_{\phi_k}f_{\bm{\phi}}(s)\,
    |\braket{s}{\varphi(\bm{\theta})}|^2,
\end{align}
which can be evaluated by computational-basis measurements of
$\ket*{\varphi(\bm{\theta})}$.

For the mixed term, only its real part is required:
\begin{align}
    \Re\!\left[K_{\theta_j\phi_k}(\bm{\lambda})\right]
    &=
    \Re\!\left[
        \bra*{\varphi(\bm{\theta})}
        \partial_{\phi_k}\hat{f}(\bm{\phi})\hat{f}(\bm{\phi})
        \partial_{\theta_j}\hat{U}
        \ket*{\bar{0}}
    \right]
    \notag\\
    &=
    \Re\left(
        \sum_{l}
        \sum_{s \in \{0,1\}^{N_q}}
        a_{j,l}\,
        \partial_{\phi_k}f_{\bm{\phi}}(s)\,
        f_{\bm{\phi}}(s)\,\right.\notag\\
    &\qquad\left.{}\times
        \braket{\varphi(\bm{\theta})}{s}\,
        \bra*{s}\hat{\mathcal U}_{j,l}\ket*{\bar 0}
    \right).
\end{align}
This quantity is evaluated using the same circuit as
Fig.~\ref{Fig: circuit}(a)-1.

The real part of $K_{\theta_j\theta_k}(\bm{\lambda})$ is written as
\begin{align}
    \Re\!\left[K_{\theta_j\theta_k}(\bm{\lambda})\right]
    &=
    \Re\left(
    \sum_{l_1,l_2}
    \sum_{s \in \{0,1\}^{N_q}}
    a^*_{j,l_1}a_{k,l_2}|f_{\bm{\phi}}(s)|^2\,\right.\notag\\
    &\qquad\left.{}\times
    \bra*{\bar{0}}\hat{\mathcal U}^{\dagger}_{j,l_1}\ket*{s}
    \bra*{s}\hat{\mathcal U}_{k,l_2}\ket*{\bar{0}}
    \right),
\end{align}
where $a^*_{j,l_1}a_{k,l_2}
=r_{j,l_1,k,l_2}e^{i\phi_{j,l_1,k,l_2}}$.
This real part can be evaluated with the circuit shown in
Fig.~\ref{Fig: circuit}(b).

% Proof correction: symmetrize the NN derivative in each bitstring pair.
We next turn to the terms involving the Hamiltonian $\hat H$.
Using Eq.~\eqref{eq: computation of a non-diagonal term with NN},
the contribution of $\hat P_\mu$ to $\Re[L_{\phi_j}(\bm\lambda)]$ is $h_\mu$ times
\begin{align}
    &\Re\!\bigl[
    \bra*{\varphi(\bm{\theta})}
    \partial_{\phi_j}\hat{f}(\bm{\phi})
    \hat P_\mu
    \ket*{\tilde{\varphi}(\bm{\lambda})}
    \bigr]
    \notag\\*
    &\quad=\frac{1}{2}
    \sum_{\substack{
        s_0=0\\
        s_{1:N_{\mathrm{q}}-1}\in\{0,1\}^{N_{\mathrm{q}}-1}
    }}
    \left(|\varphi_{+,s}|^2-|\varphi_{-,s}|^2\right)
    \notag\\*
    &\qquad\times\Big[
    \partial_{\phi_j}f_{\bm{\phi}}(s)f_{\bm{\phi}}(\tilde s)
    +f_{\bm{\phi}}(s)\partial_{\phi_j}f_{\bm{\phi}}(\tilde s)
    \Big].
\end{align}
As discussed in Sect.~\ref{sec:VQNHE}, $|\varphi_{\pm,s}|^2$ is
obtained using the measurement circuit $V$.

For the same Pauli string, the contribution to
$\Re[L_{\theta_j}(\bm\lambda)]$ is $h_\mu$ times
\begin{align}
    &\Re\!\left[
    \bra*{\partial_{\theta_j}\tilde\varphi(\bm\lambda)}
    \hat P_\mu\ket*{\tilde\varphi(\bm\lambda)}\right]
    \notag\\
    &=
    \Re\!\left[
    \bra*{\tilde{\varphi}(\bm{\lambda})}
    \hat P_\mu\hat{f}(\bm{\phi})
    \partial_{\theta_j}\hat{U}
    \ket*{\bar 0}
    \right]
    \notag\\
    &=
    \sum_{l}
    \sum_{\substack{
        s_0=0\\
        s_{1:N_{\mathrm{q}}-1}\in\{0,1\}^{N_{\mathrm{q}}-1}
    }}
    \Re\!\left[
        a_{j,l}
        \left(
        \varphi_{+,s}^{\theta_j,l}
        -
        \varphi_{-,s}^{\theta_j,l}
        \right)
    \right]\notag\\
    &\qquad{}\times f_{\bm{\phi}}(s)\,
    f_{\bm{\phi}}(\tilde s),
\end{align}
where
\begin{align}
    \varphi_{\pm,s}^{\theta_j,l}
    &:=
    \braket{\varphi(\bm{\theta})}{\pm, s_{1:N_{\mathrm{q}}-1}}\notag\\*
    &\quad{}\times
    \bra*{\pm, s_{1:N_{\mathrm{q}}-1}}\hat{\mathcal U}_{j,l}\ket*{\bar 0},
\end{align}
and $a_{j,l}=r_{j,l}e^{i\phi_{j,l}}$.
The quantity
$\Re\!\left(e^{i\phi_{j,l}}\varphi_{\pm,s}^{\theta_j,l}\right)$
is evaluated using the circuit shown in Fig.~\ref{Fig: circuit}(a)-2,
where the measurement circuit $V$ is included.

To extract the real part, the ancillary qubit is measured in the $X$-basis by applying the Hadamard gate without applying the $\hat{S}^{\dagger}$ gate. To obtain the imaginary part, the phase gate $\hat{S}^{\dagger}$ is inserted before the Hadamard gate so that the measurement basis becomes $\hat{Y}$ since $\hat S\hat{X}\hat S^{\dagger}=\hat Y$.

%%%%%%%%%%%%%%%%%%%%%%%%%%%%%%%%%%%%%%%%%%%%%%%%%%%%%%%%%%%
\section{Fit Analysis of the Fidelity Curves}
\label{app:fit_curves}
%%%%%%%%%%%%%%%%%%%%%%%%%%%%%%%%%%%%%%%%%%%%%%%%%%%%%%%%%%%
To characterize the difference between the VITE and VQNHITE fidelity curves in the main text (Figs.~\ref{fig:fidelity_nn} and~\ref{fig:fidelity_all}), we fitted the fidelity data over the interval $\beta \in [0.1,6]$.
For a given candidate fit $F^{\mathrm{fit}}(\beta)$, we define the residual at the $i$th imaginary-time point $\beta_i$ by
\begin{align}
    r_i = F_i^{\mathrm{data}} - F_i^{\mathrm{fit}},
\end{align}
where $F_i^{\mathrm{data}}$ denotes the mean fidelity at $\beta_i$ and $F_i^{\mathrm{fit}}\equiv F^{\mathrm{fit}}(\beta_i)$ is the corresponding fitted value. 
The fit parameters are obtained by minimizing the reduced chi-square:
\begin{align}
    \chi_{\mathrm{red}}^2&=\dfrac{\chi_w^2}{N_{\mathrm{data}} - N_{\mathrm{fit}}},
    \label{eq:chi2red}\\
    \chi_w^2
    &=
    \sum_{i=1}^{N_{\mathrm{data}}}
    \frac{r_i^2}{(\mathrm{SE}_i)^2},
    \label{eq:chi2w}
\end{align}
where $N_{\mathrm{data}} = 60$ is the number of fitted data points, i.e. the number of fitted imaginary-time points, $N_{\mathrm{fit}}$ is the number of fitting parameters, $\mathrm{SE}_i$ is the standard error of the mean fidelity at $\beta_i$,
\begin{align}
    \mathrm{SE}_i = \frac{\nu_i}{\sqrt{M}},
    \label{eq:SE_def}
\end{align}
with $M=100$ the number of independent runs, and $\nu_i$ is the sample standard deviation of the fidelity over the $M$ runs at $\beta_i$.

In this appendix, we tested 
two functional forms: a generalized exponential with a finite offset $F_\infty$ and a generalized inverse-power form.
The two fitting forms are
\begin{align}
    F_{\mathrm{gen}}(\beta)&= F_\infty + A\exp\!\left[-(\beta/\tau)^p\right], \label{eq:fig1_gen} \\
    F_{\mathrm{inv}}(\beta)&=\frac{1}{\left[A+(\beta/\tau)^p\right]^\alpha}, \label{eq:fig1_inv}
\end{align}
where $F_\infty$, $A$, $\tau$, $p$, and $\alpha$ are fit parameters. Thus, we set $N_{\mathrm{fit}}=4$ for both fitting forms.

We fitted both Eqs.~\eqref{eq:fig1_gen} and~\eqref{eq:fig1_inv} to all eight fidelity curves in the main text, corresponding to two ansatz connectivities (nearest-neighbor and all-to-all), two system sizes ($N=6$ and $N=8$), and two methods (VITE and VQNHITE). 
The resulting reduced chi-square values are summarized in Table~\ref{tab:chi2_compare}. 
The detailed fitting parameters for the two forms are given in Tables~\ref{tab:fit_summary_generalized} and~\ref{tab:fit_summary_inverse}.

\begin{table*}[t]
\caption{
Reduced chi-square $\chi_{\mathrm{red}}^2$ for the generalized exponential and generalized inverse-power fits. The smaller value for each dataset and method is shown in bold.
}
\label{tab:chi2_compare}
\centering
\begin{tabular}{llccc}
\toprule
Ansatz & $N$ & Method & $\chi_{\mathrm{red}}^2$ (gen) & $\chi_{\mathrm{red}}^2$ (inv) \\
\midrule
Nearest-neighbor & 6 & VITE    & 5.536 & \textbf{3.223} \\
Nearest-neighbor & 6 & VQNHITE & \textbf{1.927} & 2.017 \\
Nearest-neighbor & 8 & VITE    & 7.801 & \textbf{4.690} \\
Nearest-neighbor & 8 & VQNHITE & \textbf{1.287} & 1.476 \\
All-to-all & 6 & VITE    & 3.955 & \textbf{3.539} \\
All-to-all & 6 & VQNHITE & \textbf{1.273} & 1.294 \\
All-to-all & 8 & VITE    & 4.725 & \textbf{4.358} \\
All-to-all & 8 & VQNHITE & \textbf{1.692} & 1.693 \\
\bottomrule
\end{tabular}
\end{table*}

\begin{figure}[ht]
\centering
\includegraphics[width=\columnwidth]{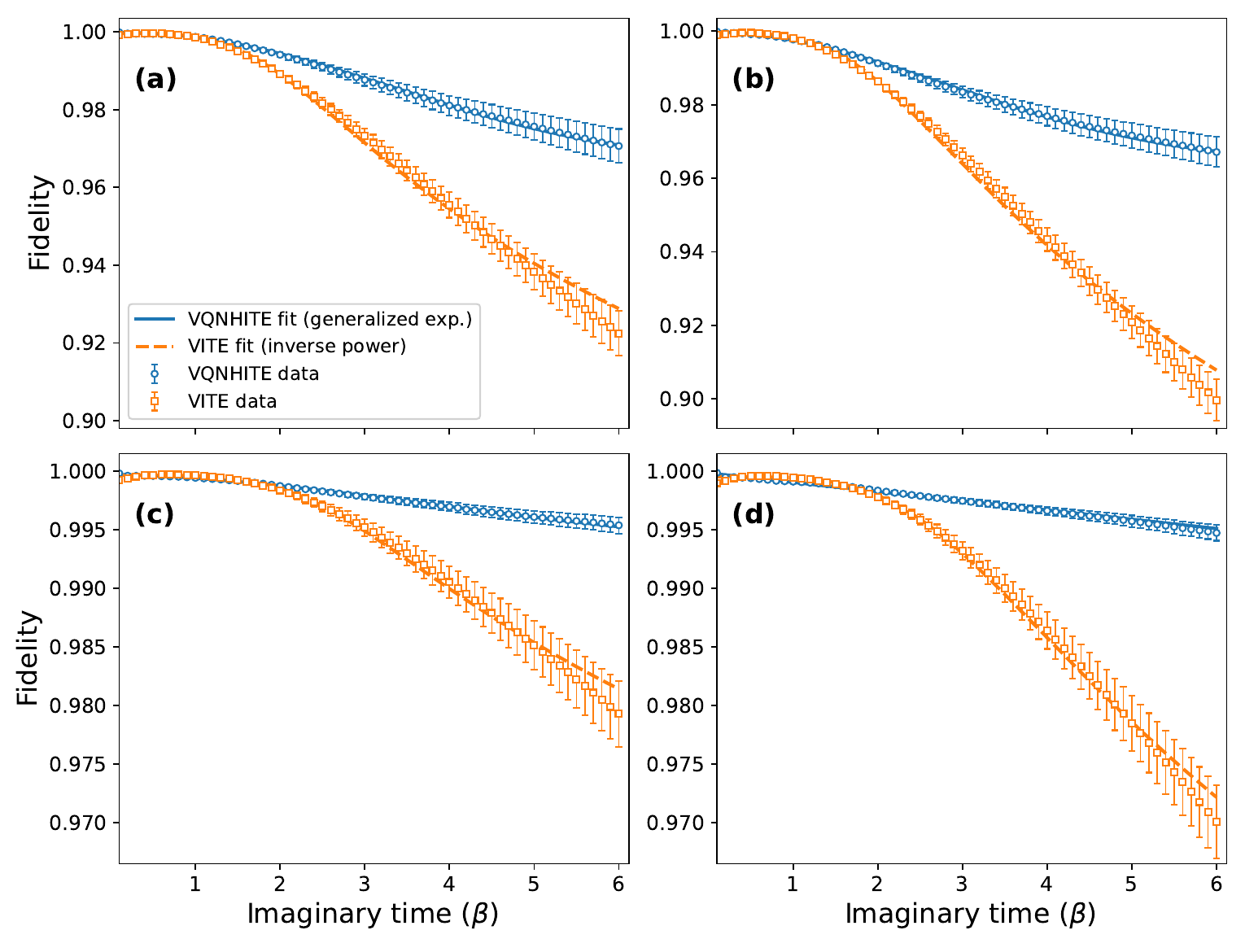}
\caption{
(Color online) Fitting results using the better of the two forms in Eqs.~\eqref{eq:fig1_gen} and~\eqref{eq:fig1_inv}, selected by the reduced chi-square in Table~\ref{tab:chi2_compare}. For all four VQNHITE curves, the selected form is the generalized exponential with a finite offset. For all four VITE curves, the selected form is the generalized inverse-power function.
Markers with error bars show the numerical data, and solid lines show the corresponding fits. Panels (a) and (b) correspond to the nearest-neighbor ansatz with $N=6$ and $N=8$, respectively, while panels (c) and (d) correspond to the all-to-all ansatz with $N=6$ and $N=8$.
}
\label{fig:generalized_exp_fits}
\end{figure}

These fitted functions should be regarded as descriptive fits within the simulated range rather than as universal asymptotic laws. 
The fitted offset $F_\infty$ should not be identified with the true $\beta \to \infty$ limit. 
The main conclusion of this analysis is that, over the range $\beta \in [0.1,6]$, VQNHITE is well described by a generalized exponential form with a finite offset for both ansatz connectivities and both system sizes, whereas VITE is better described by a non-plateaued inverse-power form, as shown in Table~\ref{tab:chi2_compare} and Fig.~\ref{fig:generalized_exp_fits}. 
This suggests that VQNHITE not only maintains a higher fidelity than conventional VITE, but also exhibits a saturation-like behavior within the simulated interval, in contrast to the continuing fidelity decay observed for VITE.

\begin{table*}[t]
\caption{
Best-fit parameters of the generalized exponential form with a finite offset $F_{\mathrm{gen}}(\beta)=F_\infty + A\exp[-(\beta/\tau)^p]$ for the fidelity curves in the main text.
}
\label{tab:fit_summary_generalized}
\centering
\begin{tabular}{llccccc}
\toprule
Ansatz & $N$ & Method & $F_\infty$ & $A$ & $\tau$ & $p$ \\
\midrule
Nearest-neighbor & 6 & VITE    & 0.935 & 0.064 &   3.584 & 3.233 \\
Nearest-neighbor & 6 & VQNHITE & 0.966 & 0.034 &   4.384 & 2.202 \\
Nearest-neighbor & 8 & VITE    & 0.914 & 0.085 &   3.698 & 3.155 \\
Nearest-neighbor & 8 & VQNHITE & 0.964 & 0.036 &   3.966 & 1.981 \\
All-to-all & 6 & VITE    & 0.982 & 0.018 &   4.188 & 3.797  \\
All-to-all & 6 & VQNHITE & 0.993 & 0.007 &   5.955 & 1.749  \\
All-to-all & 8 & VITE    & 0.972 & 0.027 &   4.373 & 3.623  \\
All-to-all & 8 & VQNHITE & 0.914 & 0.085 & 100.000 & 1.018  \\
\bottomrule
\end{tabular}
\end{table*}

\begin{table*}[t]
\caption{
Best-fit parameters of the generalized inverse-power form $F_{\mathrm{inv}}(\beta)=\left[A+(\beta/\tau)^p\right]^{-\alpha}$ for the fidelity curves in the main text.
}
\label{tab:fit_summary_inverse}
\centering
\small
\begin{tabular}{llccccc}
\toprule
Ansatz & $N$ & Method & $A$ & $\tau$ & $p$ & $\alpha$ \\
\midrule
Nearest-neighbor & 6 & VITE    & 1.021 &   2.095 & 3.944 & 0.018 \\
Nearest-neighbor & 6 & VQNHITE & 1.006 &   3.418 & 2.246 & 0.021 \\
Nearest-neighbor & 8 & VITE    & 1.020 &   2.184 & 3.817 & 0.025 \\
Nearest-neighbor & 8 & VQNHITE & 1.007 &   2.883 & 2.036 & 0.021 \\
All-to-all & 6 & VITE    & 1.069 &   2.670 & 4.594 & 0.005 \\
All-to-all & 6 & VQNHITE & 1.047 &   5.298 & 1.724 & 0.006 \\
All-to-all & 8 & VITE    & 1.051 &   2.942 & 4.197 & 0.009 \\
All-to-all & 8 & VQNHITE & 1.002 & 100.000 & 1.019 & 0.086 \\
\bottomrule
\end{tabular}
\end{table*}

\bibliographystyle{apsrev4-2}
\bibliography{arXiv_replacement}

\end{document}